\documentclass[10pt,journal,twocolumn]{IEEEtran}

\usepackage[T1]{fontenc}            
\usepackage{times}                  
\usepackage{amsmath,amssymb}        
\usepackage{graphicx}   
\graphicspath{{figures/}}

\usepackage[caption=false,font=footnotesize]{subfig} 
\usepackage[numbers,sort&compress]{natbib}  
\usepackage[colorlinks=true,citecolor=blue,linkcolor=blue,urlcolor=blue]{hyperref} 
\usepackage[margin=0.75in,columnsep=0.25in]{geometry}
\usepackage{setspace}
\usepackage{etoolbox}
\AtBeginEnvironment{thebibliography}{\fontsize{8pt}{9.5pt}\selectfont}
\makeatletter
\@ifundefined{bstctlcite}{%
  \def\bstctlcite#1{\@bsphack
    \@for\@citeb:=#1\do{%
      \edef\@citeb{\expandafter\@firstofone\@citeb}%
      \if@filesw\immediate\write\@auxout{\string\citation{\@citeb}}\fi}%
    \@esphack}
}{}
\makeatother

\makeatletter
\renewcommand\normalsize{\@setfontsize\normalsize{9}{11}}
\normalsize
\makeatother

\title{Network Models of Neurodegeneration: Bridging Neuronal Dynamics and Disease Progression}
\author{%
  Christoffer G. Alexandersen, Georgia S. Brennan, Julia K. Brynildsen,  Michael X. Henderson, Yasser Iturria-Medina, 
  and Dani S. Bassett
  \thanks{Manuscript received Month XX, 2025; revised Month YY, 2025. 
  Christoffer G. Alexandersen is with the Department of Bioengineering, University of Pennsylvania, Philadelphia, PA 19104 USA (e-mail: chrisgal@seas.upenn.edu).
  
Georgia S. Brennan is with the Institute of Molecular and Computational Medicine, Nuffield Department of Medicine, University of Oxford, Oxford OX3 9DU, U.K.; and the Mathematical Institute, University of Oxford, Oxford OX2 6GG, U.K.

Julia K. Brynildsen is with the Department of Bioengineering, University of Pennsylvania, Philadelphia, PA 19104 USA.

Michael X. Henderson is with the Department of Neurodegenerative Science, Van Andel Institute, Grand Rapids, MI 49503 USA.

Yasser Iturria-Medina is with the Department of Neurology and Neurosurgery, Montreal Neurological Institute, McGill University, Montreal, QC H3A 2B4 Canada; the McConnell Brain Imaging Centre, Montreal Neurological Institute, McGill University, Montreal, QC H3A 0G4 Canada; and the Ludmer Centre for Neuroinformatics and Mental Health, Montreal, QC, Canada.

Dani S. Bassett is with the Departments of Bioengineering, Electrical and Systems Engineering, Physics and Astronomy, Neurology, and Psychiatry, University of Pennsylvania, Philadelphia, PA 19104 USA; the Montreal Neurological Institute, McGill University, Montreal, QC H3A 0G4, Canada; and the Santa Fe Institute, Santa Fe, NM 87501 USA.}
}

\begin{document}
\maketitle
\bstctlcite{RBME:BSTcontrol}  

\begin{abstract}
Neurodegenerative diseases are characterized by the accumulation of misfolded proteins and widespread disruptions in brain function. Computational modeling has advanced our understanding of these processes, but efforts have traditionally focused on either neuronal dynamics or the underlying biological mechanisms of disease. One class of models uses neural mass and whole-brain frameworks to simulate changes in oscillations, connectivity, and network stability. A second class focuses on biological processes underlying disease progression, particularly prion-like propagation through the connectome, and glial responses and vascular mechanisms. Each modeling tradition has provided important insights, but experimental evidence shows these processes are interconnected: neuronal activity modulates protein release and clearance, while pathological burden feeds back to disrupt circuit function. Modeling these domains in isolation limits our understanding. To determine where and why disease emerges, how it spreads, and how it might be altered, we must develop integrated frameworks that capture feedback between neuronal dynamics and disease biology. In this review, we survey the two modeling approaches and highlight efforts to unify them. We argue that such integration is necessary to address key questions in neurodegeneration and to inform interventions, from targeted stimulation to control-theoretic strategies that slow progression and restore function.
\end{abstract}


\section{Introduction}

Neurodegenerative diseases, including Alzheimer's disease, Parkinson's disease, frontotemporal dementia, and amyotrophic lateral sclerosis, are devastating disorders that progressively impair brain function and ultimately lead to cognitive and motor decline. Despite their clinical differences, these diseases share a common pathological hallmark: the accumulation of misfolded protein aggregates (see~\cite{wilson_hallmarks_2023} for a review on this topic). These aggregates typically emerge in specific brain regions and then propagate along anatomical networks~\cite{liu_trans-synaptic_2012, henderson_spread_2019, luan_synaptic_2025, mcgeachan_evidence_2025, raj_network_2012}, tracking the progression of clinical symptoms and neurodegeneration~\cite{olichney_cognitive_1998, lin_plasma_2017, gomperts_amyloid_2013, bejanin_tau_2017}.

A major challenge in understanding neurodegenerative disease lies in explaining how molecular pathology gives rise to changes in brain dynamics and cognition. Neuroimaging studies using MEG, EEG, and fMRI have revealed large-scale alterations in neural communication across the course of disease~\cite{bosboom_meg_2009, luppi_oscillatory_2022, wang_altered_2006}. At the same time, cellular and \textit{in vitro} studies have identified numerous mechanisms through which protein aggregates~\cite{menkes-caspi_pathological_2015}, inflammation~\cite{odoj_vivo_2021}, and metabolic stress~\cite{sun_restoring_2025, mosconi_brain_2008} disrupt neuronal function. Together, these findings suggest a complex landscape in which macroscale dysfunction emerges from diverse microscale processes. Computational models are uniquely positioned to bridge these levels of description. By formalizing hypotheses and incorporating proposed cellular mechanisms, models can test whether these mechanisms are sufficient to account for observed changes in brain activity.

Over the past decade, a growing body of computational work has examined the effects of neurodegeneration on neuronal dynamics~\cite{stefanovski_linking_2019, bhattacharya_alpha_2011, ranasinghe_altered_2022, patow_whole-brain_2023, van_nifterick_multiscale_2022}. These efforts have yielded insight into the emergence of network dysfunction, altered oscillatory behavior, and disruptions in neuronal communication. However, these models typically assume a unidirectional flow of causality—from pathology to activity—implicitly treating pathological processes as external inputs that shape brain dynamics but are not themselves shaped by them. 

At the same time, other modeling efforts have focused on disease mechanisms, including protein spreading~\cite{raj_network_2012, iturria-medina_epidemic_2014, vogel_spread_2020, fornari_prion-like_2019} and interactions~\cite{thompson_protein-protein_2020}, glial and vascular interactions~\cite{chamberland_computational_2024, ahern_modelling_2025}, glymphatic clearance~\cite{vinje_human_2023,brennan_role_2024}, genetic regulation~\cite{sertbas_systematic_2014}, and compensatory plasticity~\cite{abuhassan_compensating_2014}. Yet these efforts also tend to omit the influence of neural activity on the disease process. This separation reflects a broader division in the field, where models of activity and models of pathology are often developed in isolation, each assuming that causality runs in one direction.

This assumption contradicts a growing body of experimental evidence showing that neuronal activity can influence disease progression. For example, neuronal activity accelerates transneuronal transport of pathological proteins~\cite{wu2016neuronal, Sokolow2015, pooler2013physiological, wu_neuronal_2020}, and neuronal stimulation has been shown to modulate—and in some cases even reverse—pathological processes~\cite{martorell_multi-sensory_2019, iaccarino_gamma_2016, suk_vibrotactile_2023}. These findings point to a fundamentally bidirectional relationship between activity and pathology: one in which dynamics and disease co-evolve and reinforce each other.

To make meaningful progress in understanding and intervening in neurodegenerative disease, we must move beyond this divide. A new generation of models is needed, ones that capture the feedback loops between neural activity and disease mechanisms across scales. Such efforts face conceptual and technical challenges, most notably the mismatch in timescales between fast neural dynamics and slow disease processes, and the absence of coarse-grained or mean-field formulations that capture their coupled interactions.

In this review, we focus on generative computational models; biologically-informed mathematical frameworks, such as systems of differential equations, that simulate brain dynamics, whether neuronal activity, prion-like spreading, or any other neurobiological process. We organize recent work into three domains. First, we survey computational models of neuronal dynamics during disease, covering approaches that aim to simulate or explain functional changes in the brain over time. Second, we review models of other disease-relevant processes, such as prion-like protein spreading, glial regulation, and clearance mechanisms. Third, we turn to the emerging and much-needed frontier of integrated modeling. Here, we discuss how combining activity and disease mechanisms can reveal critical feedback processes and may open new avenues for intervention. We emphasize the potential of such models not only to restore neural function, but also to alter the course of disease itself, and, more broadly, to reshape how we understand brain function in the face of degeneration.

\begin{figure*}[!t] 
  \centering
  \includegraphics[width=1.\linewidth]{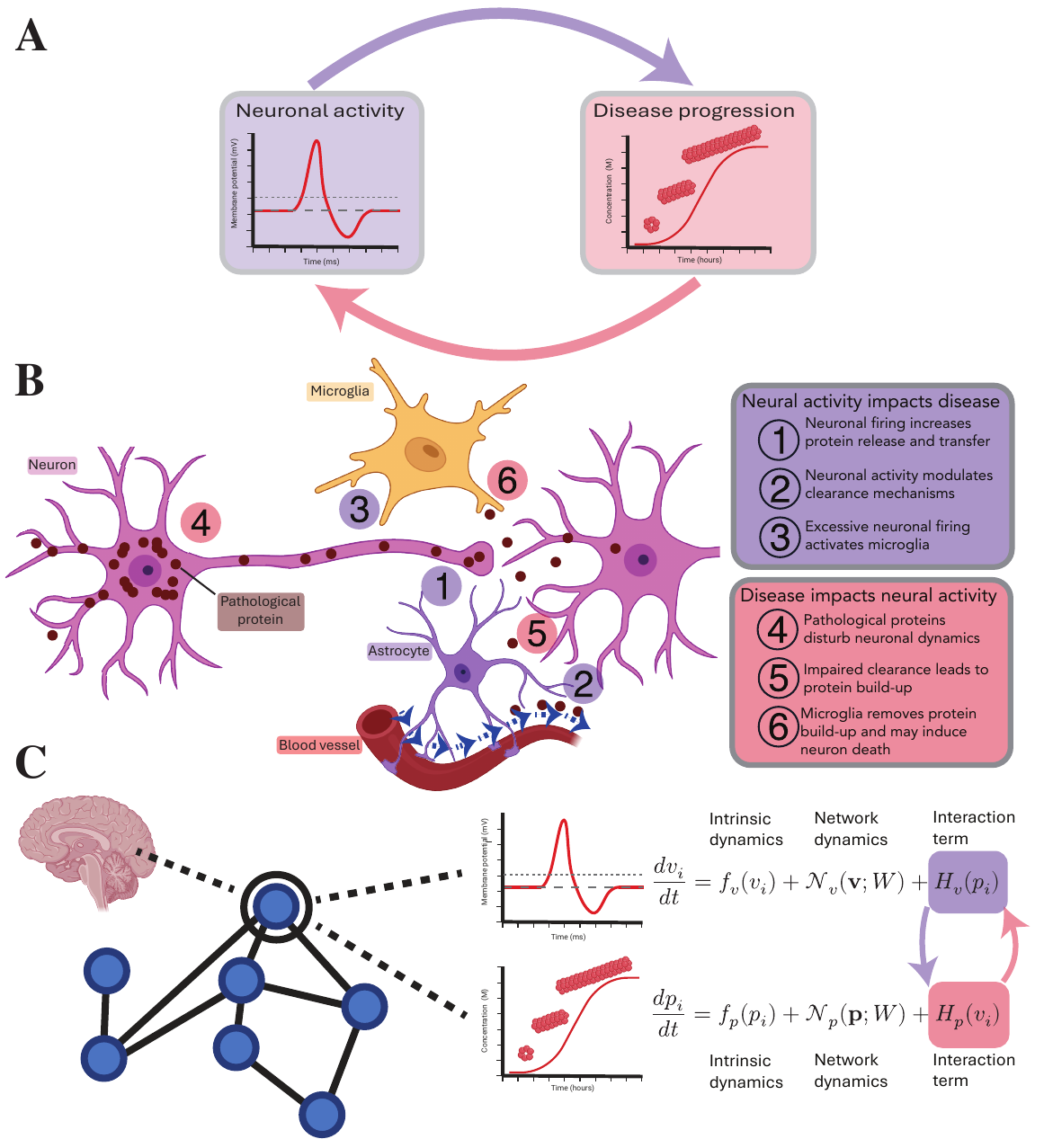}
  \caption{\textbf{Conceptual overview of activity–pathology coupling.}
\textbf{(A)} Neural activity and pathological processes form a feedback loop operating on different timescales.
\textbf{(B)} Bidirectional effects: activity can promote pathology (purple) by increasing protein release, impairing clearance, or activating glia; pathology feeds back on activity (red) by disturbing neuronal dynamics, impairing clearance, or inducing cell loss.
\textbf{(C)} Activity-spreading coupling: each region has intrinsic activity and pathology dynamics and interacts with neighbors via network coupling. Prior work typically models these processes separately; our focus is to \emph{explicitly couple} them via biologically grounded interaction terms.}
  \label{fig:Overview}
\end{figure*}

\section{Modeling Neuronal Dynamics during Neurodegenerative Disease}

Neuronal dynamics during neurodegeneration involve alterations in excitability, oscillations, and large-scale connectivity. Generative mathematical models have been developed to simulate these changes and link them to underlying mechanisms. This section first outlines key neurophysiological signatures from EEG, MEG, and fMRI, and then reviews modeling approaches across scales---from single neurons to whole-brain networks---that reproduce these signatures and identify their underlying drivers.

\begin{figure}[!t]
  \centering
  \includegraphics[width=1.\linewidth]{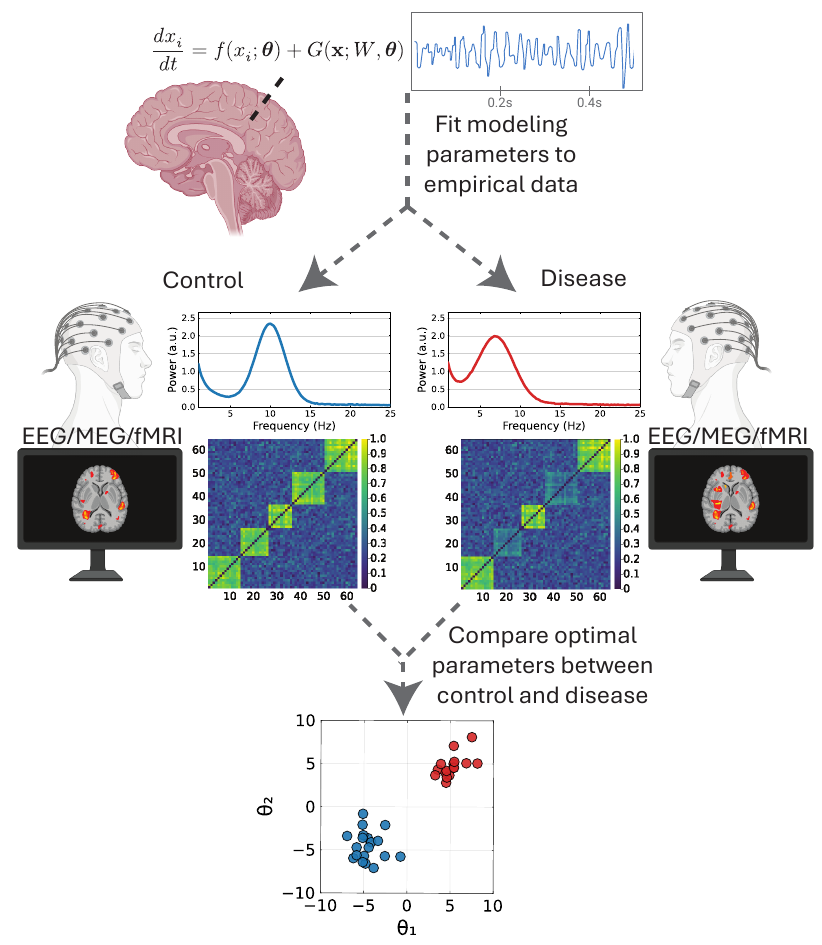}
  \caption{
\textbf{Conceptual workfllow for modeling neuronal dynamics during health and disease.}  
A whole-brain model is fit separately to control (blue) and patient (red) data. Group-level features (power spectra, functional connectivity matrices) are matched, and parameter estimates $(\theta_1,\theta_2)$ illustrate separation between groups. Differences in optimal parameters indicate specific biological properties altered in disease. 
  }
  \label{fig:neuronal_dynamics}
\end{figure}

\subsection{Neurophysiological Signatures of Dysfunction}

Neurodegenerative diseases are marked by widespread disruptions in brain dynamics. Across conditions including Alzheimer’s, Parkinson’s, frontotemporal dementia, and ALS, EEG, MEG, and fMRI studies consistently reveal alterations in large-scale neural activity. Here, we focus on neurophysiological changes relevant to modeling, rather than a comprehensive overview. These include atypical neural oscillations, altered functional connectivity, and hyperexcitability.

\subsubsection{Atypical Neural Oscillations}

A well-documented feature of Alzheimer's disease is the slowing of neural oscillations, particularly in electrophysiological recordings. E/MEG studies have consistently reported slower dominant alpha rhythms, along with reduced alpha power alongside increased delta and theta activity~\cite{hsiao_altered_2013, moretti_individual_2004, wang_multiple_2015, gouw_eeg_2017}. These spectral changes reflect a shift in the brain’s dominant rhythms and are thought to be related to cognitive decline~\cite{bruna_meg_2023, engels_slowing_2016}.
Similar patterns of oscillatory slowing have also been observed in Parkinson's disease~\cite{olde_dubbelink_cognitive_2013}, suggesting that this phenomenon may be a general signature of network-level disruption across neurodegenerative diseases.

In Parkinson's patients, symptoms of bradykinesia/akinesia and rigidity have been shown to be related to enchanced beta oscillations in basal ganglia nuclei~\cite{ray_local_2008, kuhn_high-frequency_2008}, and surpression of beta oscillations in this region improves motor symptoms~\cite{kuhn_high-frequency_2008}. More recent evidence distinguishes between low‑beta (13–20 Hz) and high‑beta (20–35 Hz) sub-bands in subthalamic nucleus activity. High‑beta power reliably predicts motor improvement from deep brain stimulation, accounting for around 37\% of variance in bradykinesia–rigidity outcomes, whereas low‑beta does not show such predictive value~\cite{chen_subthalamic_2022}. Thus, discrete assessment of these sub‑bands may enhance the specificity of electrophysiological biomarkers for monitoring disease state and guiding adaptive deep brain stimulation.

\subsubsection{Alterations in Functional Connectivity}

Changes in interregional communication are another hallmark of neurodegeneration. In Alzheimer's disease, M/EEG studies have revealed impaired functional connectivity in the alpha and beta bands, particularly affecting hub regions in the posterior default-mode network~\cite{stam_graph_2009}. Individuals with Parkinson's disease also show frequency-dependent connectivity changes, including increased theta, alpha, and beta synchronization that correlates with motor symptoms~\cite{stoffers_dopaminergic_2008}. fMRI provides a complementary view, revealing both increases and decreases in large-scale connectivity over the course of disease progression. In early Alzheimer's disease, connectivity within anterior and ventral components of the default-mode network is often elevated, followed by widespread decreases as the disease advances~\cite{damoiseaux_functional_2012}. Parkinson's disease similarly exhibits a mixed pattern: akinetic Parkinson's is associated with reduced frontal connectivity, while motor-related regions may show increased coupling~\cite{sabatini_cortical_2000, baudrexel_resting_2011}.

\subsubsection{Hyperexcitability and Epileptiform Activity}

An emerging theme in both animal models and human studies is the presence of hyperexcitability and epileptiform activity during early stages of neurodegeneration. In AD mouse models, early tau and amyloid pathology is associated with increased neuronal firing and network hypersynchrony~\cite{busche_critical_2012}. Supporting this relationship, MEG studies in patients have detected signs of early hyperactivity~\cite{koelewijn_oscillatory_2019}, and epidemiological data show increased incidence of seizures and subclinical epileptiform discharges in Alzheimer's disease~\cite{vossel_incidence_2016, vossel_seizures_2013}. Although hyperexcitability has received less attention in Parkinson’s disease, findings from both pathology-induced and genetic mouse models report hyperexcitable neurons in the motor cortex and hippocampus~\cite{chen_motor_2025, tweedy_hippocampal_2021}, suggesting that circuit-level instabilities may generalize across disease types.

\subsection{Using Computational Models of Neuronal Activity to Understand Functional Decline}

Computational models are widely used to investigate how neurodegenerative disease alters brain activity. These models span spatial scales from individual neurons to whole-brain networks. At each scale, they explain changes in excitability, oscillations, and connectivity, linking biological mechanisms to EEG, MEG, and fMRI observations. These studies typically identify parameter regimes that reproduce disease phenomena and compare them with those producing healthy dynamics (Figure~\ref{fig:neuronal_dynamics}). Altered parameters are interpreted as candidate mechanisms for pathological changes. Some models instead capture general trends such as oscillatory slowing or hyperactivity without direct data comparison.

\subsubsection{Single-Neuron Models}
Single-neuron models aim to reproduce the electrophysiological properties of individual neurons, such as how they generate action potentials and respond to inputs. They are the most fine-grained of the modeling approaches we discuss, focusing on the dynamics of a single cell.
These models span a wide range of complexity, from detailed Hodgkin–Huxley–type descriptions with multiple ion channels to reduced firing-rate or quadratic integrate-and-fire (QIF) formulations. A generic conductance-based form for the membrane potential $V_i$ for neuron $i$ is
\begin{equation}
C \frac{dV_i}{dt} = -I_{\mathrm{ion}}(V_i,g_i) + \sum_j w_{ij} S_j(t) + I_\mathrm{ext}(t),
\end{equation}
where $I_{\mathrm{ion}}$ are ionic currents parameterized by conductances $g$, the summation captures synaptic inputs with weights $w_{ij}$, $C$ is the membrane capacity, and $I_\mathrm{ext}$ represents external drive.
This framework allows researchers to investigate how alterations in intrinsic excitability, synaptic input, or ion channel function might explain disease-related changes in neuronal behavior.

Relatively few studies have applied such models directly to neurodegeneration. One study of hippocampal CA1 neurons in Alzheimer's disease showed that disturbed excitation-inhibition balance, ion channel dysregulation, and increased excitatory drive can jointly account for the hyperexcitability observed in early disease stages~\cite{mittag_modelling_2023}. Another model explored how synaptic degradation in CA1 pyramidal neurons could lead to reductions in alpha power, consistent with EEG observations~\cite{dong_how_2022}. Using a Hodgkin-Huxley model for hippocampal neurons, acetylcholine deficiency and amyloid-$\beta$ pathology were both capable of decreasing firing rates and increasing relative delta power~\cite{jiang_dynamics_2020}. Another Hodgkin-Huxley model formalism was used to explain experimentally-observed changes in an Alzheimer's mouse model, where increased sodium leak and HCN channel conductance replicated observed action potential dynamics in dentate gyrus interneurons, but could not characterize changes in membrane potential~\cite{perez_analyzing_2016}.
A recent study used the analytically tractable firing rate distributions of Gauss--Rice neurons, fitted to Neuropixels data, to show that amyloid precursor protein (APP) modulates cortico-hippocampal activity via NMDA receptor interactions~\cite{harris_amyloid_2025, schmidt_analytically_2024}. Although these models offer biologically grounded insight into local mechanisms, they typically include a high number of parameters and are difficult to analyze when compared to models such as neural masses. One particular downside arising from the high-dimensionality in parameter space is functional degeneracy, where pathological changes in neuronal activity cannot be attributed to a single mechanism in the model. 

\subsubsection{Neural Mass Models}

Neural mass models describe the average activity of neuronal populations, often in terms of mean membrane potential or firing rate. Traditional formulations (e.g., Wilson–Cowan, Jansen–Rit) are based on assumed input–output relations and synaptic dynamics at the population level~\cite{wilson_excitatory_1972, jansen_electroencephalogram_1995}, while more recent work derives neural mass models directly from spiking neurons such as the QIF~\cite{montbrio_macroscopic_2015}; a powerful example is the Ott–Antonsen ansatz, which provides an exact reduction in the limit of infinitely large populations~\cite{ott_low_2008}.
They are particularly suited to studying collective oscillations and have been widely applied to EEG and MEG data.
A canonical example is the Wilson–Cowan framework, which couples the average firing rates of excitatory ($E$) and inhibitory ($I$) populations in a way that generates oscillations via a Hopf bifurcation:
\begin{align}
\tau_E \frac{dE}{dt} &= -E + \phi(w_{EE}E - w_{EI}I + I_E), \\
\tau_I \frac{dI}{dt} &= -I + \phi(w_{IE}E - w_{II}I + I_I),
\label{eq:neural-mass}
\end{align}
where $\phi$ is a nonlinear transfer function, $w_{XY}$ are synaptic coupling strengths, $\tau_{E \backslash I}$ are timescale parameters, and $I_E, I_I$ are external inputs.
This excitatory–inhibitory loop provides a canonical mechanism for generating neural oscillations.  Other neural mass models may include more neuronal types (e.g., Jansen–Rit), stochastic drive, or delayed coupling between regions.
Because of their tractability, neural mass models have been used to explore oscillatory slowing in Alzheimer’s disease and enhanced beta oscillations in Parkinson’s disease. More generally, neural mass models are coarse-grained and biologically simplified, which limits their ability to pinpoint precise mechanisms underlying pathological changes.

Oscillatory slowing, in particular slowing of the resting state alpha rhythm with concurrent increases in theta rhythms, is a hallmark of Alzheimer's disease neurophysiology. A thalamic model incorporating increased inhibition from thalamic reticular nucleus neurons reproduced the alpha and theta atypicalities observed in Alzheimer's patients~\cite{bhattacharya_alpha_2011, bhattacharya_thalamocorticothalamic_2011}.
Another study instead linked cholinergic depletion to oscillatory slowing of the alpha rhythm in Alzheimer's~\cite{yang_effect_2022}. Related work has shown that Jansen–Rit models can also reproduce dynamic fluctuations in alpha power and their relationship to functional connectivity~\cite{cabrera-alvarez_fluctuations_2025}. A recent study used a laminar neural mass model incorporating excitatory populations and fast‑spiking parvalbumin (PV) interneurons to simulate progressive PV dysfunction (via reduced PV-to-pyramidal connectivity) and later pyramidal neuron impairment to reproduce Alzheimer's-related electrophysiological biomarkers, showing early-stage hyperexcitability with increased gamma/alpha power, transitioning to oscillatory slowing and hypoactivity in later stages~\cite{sanchez-todo_fast_2025}. This modeling supports the idea that local PV interneuron dysfunction drives early oscillatory atypicalities, whereas pyramidal cell loss underlies later spectral decline and reduced firing.

As for the enhanced beta oscillations observed in the basal ganglia in Parkinson’s patients, a model of the reciprocally connected subthalamic nucleus (STN) and globus pallidus externus (GPe) showed that Parkinsonian beta oscillations emerge from strengthened STN–GPe coupling and elevated cortical drive~\cite{holgado_conditions_2010}, which was also replicated in another study where a neural mass model of the basal ganglia circuit showed elevated beta oscillations with stronger STN-GPe coupling and intrinsic neuronal excitability~\cite{liu_neural_2016}. Dynamic causal modeling of LFP data in a rodent Parkinson’s model revealed that increased cortex-to-STN drive and altered STN–GPe inhibition underlie exaggerated beta synchrony~\cite{moran_alterations_2011}. 
Moreover, a model of the basal ganglia--thalamocortical loop identified self-inhibition of the globus pallidus externus to modulate whether changes occur in upper- and lower-band beta oscillations~\cite{liu_neural_2017}, where upper band changes are more strongly associated with motor symptoms~\cite{chen_subthalamic_2022}.

In both Alzheimer’s and Parkinson’s disease, progressive neuronal loss accompanies the oscillatory changes captured by neural mass models. While most neural mass models mimic neuronal loss indirectly, by reducing synaptic connectivity parameters, this approach may not accurately capture the effects of neuronal loss. A recent extension of the neural field framework incorporated spatially variable neuron density to explicitly represent neuronal loss, which resulted in altered synchronization patterns and oscillatory dynamics~\cite{reyes_modeling_2022}. This provides a theoretical link between the spatial progression of neurodegeneration and the emergence of population-level activity changes, complementing the more phenomenological approaches used in traditional neural mass models.

A wealth of empirical studies have demonstrated altered functional connectivity in neurodegenerative disease as measured by E/MEG and fMRI. Neural mass models may only capture local changes in neural oscillations, not correlations in neural activity between regions of interest. However, neural mass models may be connected in a network given by whole-brain structural connectivity and used to study functional connectivity. These whole-brain models of neural dynamics have recently provided important insights into the deviations in functional connectivity associated with neurodegenerative disease.

\subsubsection{Whole-Brain Models}

Whole-brain models are built by assigning a neural mass model to each brain region and coupling them according to empirically derived structural connectivity, typically from diffusion MRI. In this way, local population dynamics become the nodes of a network, while long-range anatomical pathways define the edges.
A common example uses Wilson–Cowan–type neural masses \eqref{eq:neural-mass}, where each region $i$ is represented by excitatory ($E_i$) and inhibitory ($I_i$) populations coupled through the connectome:
\begin{align}
\tau_E \frac{dE_i}{dt} &= -E_i + \phi\left(w_{EE}E_i - w_{EI}I_i + G\sum_j W_{ij}E_j + I_{E,i}\right), \\
\tau_I \frac{dI_i}{dt} &= -I_i + \phi\left(w_{IE}E_i - w_{II}I_i + I_{I,i}\right),
\end{align}
where $W_{ij}$ is the structural connectivity matrix representing long-range neuronal connections, $G$ a global coupling parameter, and $\phi$ a nonlinear transfer function. Long-range coupling is typically modeled as excitatory-to-excitatory input only, though this may vary.
This formulation illustrates how excitatory–inhibitory population dynamics can be embedded in a large-scale network. The choice of local neural mass is flexible—Wilson–Cowan, Jansen–Rit, reduced Wong–Wang, or other formulations may be used depending on the application. Interregional coupling can also incorporate stochasticity, conduction delays, or heterogeneous regional parameters to better reflect biological variability.
These models provide a tractable framework for studying how interactions between brain regions give rise to emergent large-scale phenomena such as functional connectivity, network oscillations, and other neurophysiological changes observed in health and disease. In practice, whole-brain models are often parameterized so that their simulated functional connectivity best reproduces empirical FC from neuroimaging data, enabling direct comparison between model predictions and experimental observations, as illustrated in Figure~\ref{fig:neuronal_dynamics}.

Several studies have used whole-brain models to examine how neurodegenerative pathology affects global dynamics. One attractive approach is to use empirical measures of protein pathology such as PET imaging to quantify the amount and type of protein in each region of interest, and to subsequently use this information to infer its impact on neuronal dynamics.
Stefanovski \textit{et al.}~\cite{stefanovski_linking_2019} used such an approach to study aberrant neuronal dynamics observed in Alzheimer's disease, hypothesizing that amyloid-$\beta$ disrupts the activity of inhibitory neurons in cortical regions. Utilizing amyloid-$\beta$ PET data to modulate local inhibitory gain in a Jansen--Rit whole-brain model, the authors reproduced the oscillatory slowing of the alpha rhythm and concurrent increases in theta power as observed in EEG for Alzheimer's patients. Subsequent studies have extended this approach by not only incorporating protein neuroimaging, but also functional neuroimaging data such as fMRI and E/MEG. In this approach, empirical protein pathology maps inform regional model parameters, and functional data are used to infer how these changes alter neuronal dynamics to match observed neuroimaging patterns. 

Building on this combined protein-functional approach, Patow \textit{et al.} (2023)~\cite{patow_whole-brain_2023} used a whole-brain model of reduced Wong-Wang neural masses combined with fMRI along with both amyloid-$\beta$ and tau PET. The authors assumed that amyloid-$\beta$ reduces inhibitory neuron activity and tau reduces excitatory neuron activity, with the magnitudes of these effects inferred by fitting the computational model to resting-state fMRI data from individuals with Alzheimer's disease. The authors demonstrated that including the PET protein data and its impact on neuronal dynamics improved the whole-brain model's fit to functional connectivity, and that the effects of amyloid-$\beta$ are more apparent in early-stage Alzheimer's which are then later dominated by the effects of tau pathology. 
In a related study, Ranasinghe \textit{et al.} (2022)~\cite{ranasinghe_altered_2022} used a similar approach, but modeled each brain region independently as an isolated neural mass fitted to recapitulate MEG power spectra in Alzheimer's patients. The optimal neural mass parameters fitted to Alzheimer's MEG were found to be associated with regional protein levels of amyloid-$\beta$ and tau. These findings demonstrated that tau PET correlated with excitatory neuron parameters whereas amyloid-$\beta$ PET correlated with inhibitory neuron parameters.

Extending this line of work toward individualized modeling, Sanchez-Rodriguez \textit{et al.} (2024)~\cite{sanchez-rodriguez_personalized_2024} used resting-state fMRI along with individual amyloid-$\beta$ and tau PET scans to infer the impact of protein pathology on neuronal excitability in the Alzheimer's disease spectrum using a whole-brain modeling approach. For each participant, a whole-brain model of Wilson-Cowan neural masses was fitted to optimally reconstruct fMRI resting-state indicators, but where regional amyloid and tau levels were allowed to affect the excitablity of each region (affecting both excitatory and inhibitory populations equally), without any assumption about whether the protein increases or decreases excitability. For most Alzheimer's participants, amyloid-$\beta$ was inferred to increase hyperexcitability, and the amount of this increase was associated with worse cognitive performance. As for tau levels, results were more mixed between patients, but were generally associated with increased excitability. The data-derived excitability values correlated with clinically relevant AD plasma biomarker concentrations (e.g., p-tau217, p-tau231) and grey matter atrophy, while also reproduced oscillatory slowing as observed in E/MEG. A subsequent model application ~\cite{sanchez-rodriguez_-vivo_2024} used fMRI-derived excitability values in combination with neurotypical brain transcriptomes (Allen Human Brain Atlas) to identify molecular mechanisms spatially associated with neuronal dysfunction in AD, and to rank pharmacological candidates according with their potential to modify AD progression.

Whereas these studies explicitly linked protein pathology to neuronal excitability, other work has instead concentrated on functional alterations captured by neuroimaging modalities alone. For example, Van Nifterick \textit{et al.} (2022)~\cite{van_nifterick_multiscale_2022} used a whole-brain model to test literature-based hypotheses for the early-stage hyperexcitability observed in Alzheimer's, independently testing six mechanisms of pyramidal hyperexcitability and inhibitory neuronal dysfunction. Moreover, whole-brain simulations were compared to empirical MEG spectra, verifying that five out of six putative mechanisms exhibited both hyperexcitability \emph{and} oscillatory slowing, demonstrating that these two phenomena may arise from the same mechanism which may result from either hyperexcitable pyramidal neurons or dysfunctional inhibitory neurons. 

Zimmermann \textit{et al.} (2018)~\cite{zimmermann_differentiation_2018} built personalized large-scale brain models of reduced Wong-Wang neural masses (using The Virtual Brain platform) for 124 individuals spanning healthy aging, mild cognitive impairment, and Alzheimer's. By adjusting global coupling, conduction velocity, and within-region excitation-inhibition balance parameters for each subject, they could accurately simulate that individual’s resting-state fMRI connectivity. Notably, the fitted model parameters correlated strongly with cognitive performance and distinguished Alzheimer's from controls better than structural connectivity alone; however, the optimal parameters found for individuals did not differ significantly between the control and clinical groups. 

A similar approach examines how functional connectivity evolves over the course of Alzheimer’s disease, as in Demirta\c{s} \textit{et al.} (2017)~\cite{demirtas_whole-brain_2017}, who modeled whole-brain dynamics using Hopf oscillators. In particular, they fit the whole-brain model to healthy control resting-state fRMI functional connectivity, and then postulated that globally decreasing the Hopf bifurcation parameter in each region (which decreases oscillation amplitudes and eventually stops oscillations altogether) will better reproduce the functional connectivity of Alzheimer's, which they confirm where the optimal fit for the advancing stages are found for lower values of the bifurcation parameter. 

More recent work has combined whole-brain models with machine learning approaches. 
Sanz-Perl \textit{et al.} (2023)~\cite{sanz_perl_model-based_2023} used whole-brain modeling of Hopf oscillators combined with an autoencoder approach to study the impact of perturbations on whole-brain dynamics in Alzheimer's disease and behavioural variant frontotemporal dementia (bvFTD). Optimizing simulation fit to fMRI resting-state for controls, AD, and bvFTD, they found that using anamotical priors such as Alzheimer's and bvFTD atrophy maps to infer region-specific parameters resulted in better fits for the clinical groups. Moreover, differences in optimal parameter fits showed significant differences in hippocampal-specific parameters in Alzheimer's (shifted towards lower activity levels without oscillatory dynamics) and in the bilateral insula (shifted towards higher activity with oscillatory dynamics) in bvFTD. Further, encoding the simulated functional connectivity for the different clinical groups using a variational autoencoder into a two-dimensional representation, a perturbational landscape was carved out investigating different stimulation protocols and their effectiveness in pushing the diseased representational state toward the healthy, control state. In particular, they found that stimulation of the temporal lobe, and the hippocampus in particular, pushed Alzheimer's states toward the healthy control state, while frontal regions were more effective for bvFTD. 
 
Other approaches have used neural mass models parameterized to have up- and down-states, as opposed oscillatory behaviour. Yalçınkaya \textit{et al.} (2023)~\cite{yalcinkaya_personalized_2023} used a whole-brain model of an exact mean-field reduction of QIF neuronal populations~\cite{montbrio_macroscopic_2015} with fMRI data to capture changes in homotopic connections and limbic network dynamical fluidity. The QIF mean-field reduction model was parameterized in a bistable regime with two stable fixed points of high (up-state) and low (down-state) firing rates. The global coupling coefficient and excitability in the limbic network were inferred using simulation-based inference, and showed differences between healthy and Alzheimer's patients. 

Subsequent studies have broadened the scope of biological mechanisms incorporated in the modeling approach. For example,
Depannemeacker \textit{et al.} (2025)~\cite{depannemaecker_next_2025} added neuromodulatory dynamics, such as that of dopamine re-uptake crucial in Parkinson's, to the QIF exact mean-field reduction with adaptive and conductance-based dynamics~\cite{chen_exact_2022, sheheitli_incorporating_2024}. 
Angiolelli \textit{et al.} (2025)~\cite{angiolelli_virtual_2025} used this exact mean-field neural mass in a whole-brain framework to correctly infer what EEG and deep electrode recordings belong to Parkinson's patients with or without L-Dopa administration, showing that whole-brain modeling can capture changes in whole-brain dynamics associated with heightened dopaminergic tone. 

These models offer a powerful framework for linking molecular pathology to network-level dysfunction. Nonetheless, they do not aim to explain the progression or emergence of pathology, which may be more central to informing new treatment strategies. Moreover, most of them implicitly assume a unidirectional flow of causality from pathology to neural activity. This assumption is at odds with a growing body of experimental evidence demonstrating that neural activity can itself modulate the production, spread, and clearance of pathological proteins. Capturing this bidirectional interplay will require integrated approaches that couple biological mechanisms with neuronal dynamics. At the same time, whole-brain models face practical challenges, including parameter redundancy, limited experimental validation, and interpretational variability across modeling choices, which constrain their broader applicability.

\subsection{Limitations of Modeling Neuronal Activity in Isolation}
Models of neuronal dynamics have provided valuable insight into how neurodegenerative disease alters brain function, explaining phenomena such as hyperactivity, oscillatory slowing, and altered functional connectivity. They have also offered plausible mechanisms for specific observations, including enhanced beta oscillations in Parkinson's disease, slowing of the resting-state alpha rhythm in Alzheimer's disease, and the influence of amyloid-$\beta$ and tau on excitatory–inhibitory balance. However, these models typically treat pathology as an external input, excluding it from the dynamical equations. This choice leaves key questions unanswered about the origin, spatial distribution, and progression of neurodegenerative pathology.

Addressing these questions requires going beyond neural activity. Many hallmark features of neurodegeneration---such as the stereotyped onset of pathology in specific regions and its propagation through connected networks---depend on biological processes not represented in neuronal activity models, including protein misfolding, prion-like spreading, regional vulnerability, gene expression, and clearance dynamics.

Computational models of disease progression, particularly those describing the transneuronal spread of misfolded proteins, have provided important insights into these processes. In the next section, we review this class of models, along with others that simulate glial, vascular, and metabolic contributions to disease. While these approaches complement activity-based models by explaining where and how pathology unfolds, they too often neglect the influence of neuronal dynamics on disease evolution, a gap that integrated modeling seeks to close.

\section{Modeling Disease Progression and Other Biological Processes}

Neurodegenerative diseases involve many biological processes, but a common feature is abnormal protein aggregation~\cite{jucker2013self,frost2010prion, brettschneider_spreading_2015}. Each disease is associated with hallmark proteins, for example, tau and amyloid-$\beta$ in Alzheimer’s disease or $\alpha$-synuclein in Parkinson’s disease, though these pathologies can also overlap across conditions. These proteins are thought to mediate neuronal dysfunction and propagate from one neuron to another along axonal connections~\cite{frost2009propagation,de2012propagation}. This supports the hypothesis that disease progression follows anatomical connections, giving rise to spatial patterns such as Braak staging~\cite{braak1991neuropathological}.

\begin{figure}[!t]
    \centering
    \includegraphics[width=\linewidth]{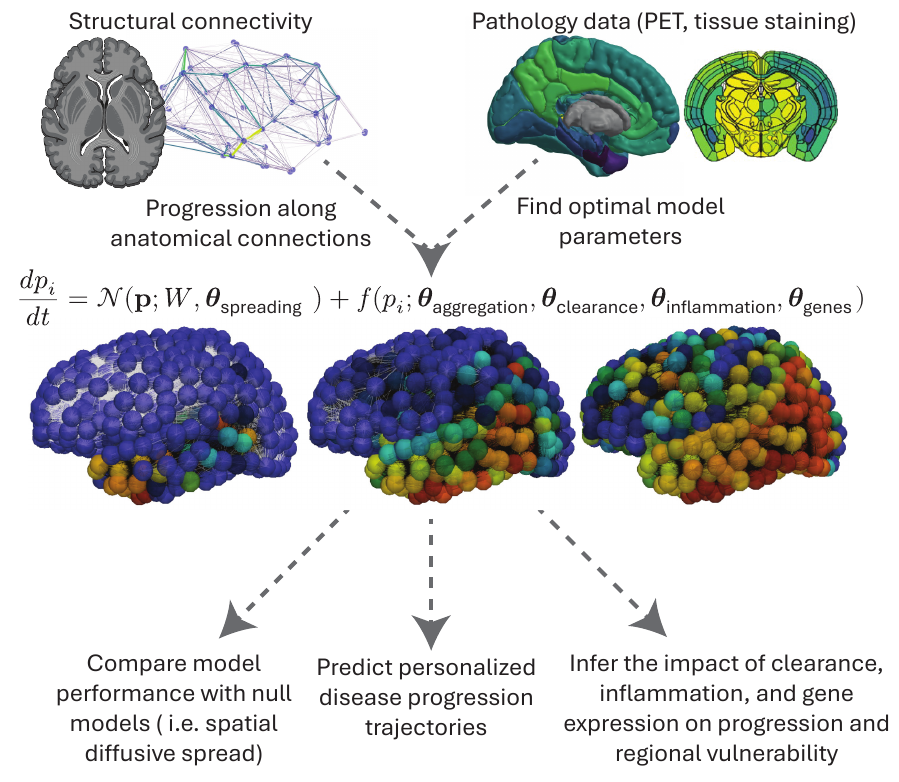}
    \caption{
        \textbf{Conceptual workflow for modeling prion-like protein spreading in neurodegeneration.} 
Progression models of protein pathology simulate disease spread along the structural connectome derived from diffusion MRI. Model parameters are fit to empirical pathology (e.g., PET or histology) and evaluated against spatial/diffusion nulls. The calibrated model can address scientific questions such as whether connectivity alone explains progression, which regions show excess vulnerability, where initiation likely occurred, and how mechanisms like clearance, inflammation, or gene regulation shape trajectories.
    }
    \label{fig:workflow}
\end{figure}

To test this hypothesis, researchers have developed computational models that simulate how protein pathology spreads through the brain. These models are designed not only to reproduce observed patterns of progression but also to assess how well spreading along anatomical connections explains disease trajectories in humans. Among them, simple network-based models have been especially productive. These models describe how protein pathology progresses throughout the brain as a system of differential equations, and can be calibrated to pathology data to test biological hypotheses, as illustrated in Figure~\ref{fig:workflow}. Other approaches extend these models to include processes such as active transport, inflammation, metabolism, and clearance. The following sections examine each of these processes in turn, highlighting their distinct contributions and interactions within spreading models.

\subsection{Models of Prion-like Spreading}

A major class of disease progression models centers on the hypothesis that pathological proteins propagate along anatomical connections. The brain’s structural organisation as a small-world network~\cite{bassett2006small} naturally motivates a network-based modelling approach. The simplest such models assume passive diffusion along axonal connections, whereby pathology spreads through the brain’s structural network. This process is typically captured by a linear diffusion equation:
\begin{equation}
\frac{du}{dt} = -\rho L u,
\end{equation}
where $u$ is the vector of regional pathology levels, $L$ is the graph Laplacian of the structural connectome, and $\rho$ is a diffusion rate constant~\cite{raj_network_2012, raj_network_2021, freeze_network_2020}. 

More biologically grounded models add local protein conversion dynamics. In prion-like processes, misfolded proteins recruit healthy ones and convert them into the same pathological form~\cite{fornari_prion-like_2019}. This has been formalized using epidemic-style spreading models~\cite{iturria-medina_epidemic_2014} and through the Smoluchowski equations, which describe protein aggregation kinetics in terms of nucleation, growth, and fragmentation~\cite{fornari_spatially-extended_2020}. Extensions of this framework have incorporated clearance~\cite{meislclearance2020}, therapeutic antibody effects~\cite{linse2020kinetic}, and even brain-scale modeling of treatment strategies~\cite{brennan2025network}. Because the full Smoluchowski framework is mathematically complex, simplified models are often used instead. One such example is the heterodimer model, which can be understood as a two-species truncation of the Smoluchowski equations. It describes the dynamics of healthy ($u$) and misfolded ($v$) variants of the same protein:
\begin{align}
\frac{du}{dt} &= -\rho L u + k_0 - k_1 u - k_2 u v, \\
\frac{dv}{dt} &= -\rho L u - k_3 v + k_2 u v.
\end{align}
This captures replication, clearance, and the conversion from healthy to misfolded states~\cite{weickenmeier_physics-based_2019,borgqvist_hemito-dynamics_2025}, and can be extended to account for interactions between multiple misfolded variants~\cite{thompson_protein-protein_2020}.
An even more tractable alternative is the Fisher–KPP equation, which can be derived as a further simplification of the heterodimer model~\cite{fornari_prion-like_2019, putra2023front}:
\begin{equation}
\frac{du}{dt} = -\rho L u + \alpha u (\beta - u).
\end{equation}
This form combines diffusive spread with local growth, while requiring only a few parameters. Its tractability makes it especially useful for inference and predictive modeling, while still retaining the key features of prion-like transmission. Beyond biology, the same formalism has also inspired physics-based analogues, where mechanical metamaterials mimic prion-like propagation of conformational states~\cite{ouellet_mechanical_2025}.

Computational models of neurodegenerative disease have provided evidence that pathological proteins such as tau and $\alpha$-synuclein spread through the brain along anatomical connections. In Alzheimer’s disease, early work showed that spreading along structural connections correlates with regional atrophy~\cite{raj_network_2012}. Fisher–KPP models reproduce Braak staging~\cite{putra2021braiding}, and stochastic epidemic-spreading models explain the spatiotemporal distribution of amyloid PET~\cite{iturria-medina_epidemic_2014} and the variance of tau PET~\cite{vogel_spread_2020}, while also suggesting that amyloid accumulation is driven primarily by clearance deficiency rather than overproduction~\cite{iturria-medina_epidemic_2014, brennan_role_2024}. Tau PET covariance patterns further align with intrinsic functional networks~\cite{ossenkoppele_tau_2019}, and network diffusion models have been shown to reproduce longitudinal tau PET progression~\cite{schafer_network_2020}. In vivo kinetic analyses indicate that tau accumulation is limited mainly by slow local replication rather than fast interregional spread, with tau seeds doubling only every five years~\cite{meisl_vivo_2021}. Other network-based studies show that tau and microglia follow gradients of structural and functional connectivity~\cite{ottoy_tau_2024}, that APOE and glutamatergic gene-expression gradients shape network vulnerability~\cite{montal_network_2022}, and that interactions between amyloid-$\beta$ and tau at specific hubs accelerate propagation~\cite{lee_regional_2022}. Earlier symptom onset is linked to stronger tau burden in globally connected hubs, resulting in faster spreading and cognitive decline~\cite{frontzkowski_earlier_2022}. Because Fisher–KPP-style models require only a few parameters, they are also well suited for Bayesian inference and have been used to accurately predict longitudinal tau PET trajectories~\cite{schafer_predicting_2021, schafer_correlating_2022, chaggar_personalised_2025}. Beyond hypothesis-driven approaches, machine learning methods combining physics-informed neural networks and symbolic regression have even discovered reaction–diffusion equations directly from longitudinal tau PET data~\cite{zhang_discovering_2024}. At the same time, evidence from mouse models suggests that amyloid-$\beta$ spread may be driven more by extracellular proximity than by connectivity~\cite{mezias_analysis_2017}.
In primary tauopathies, PET and post-mortem data likewise demonstrate that tau deposition patterns follow functional connectivity, particularly for neuronal compared to glial tau, supporting connectivity-mediated spread in 4R tau disorders~\cite{franzmeier_tau_2022}.

In Parkinson’s disease, diffusion and epidemic spreading models reproduce the progression of atrophy and $\alpha$-synuclein pathology. Linear diffusion models correlate with longitudinal pathology observed in mouse models~\cite{henderson_spread_2019, cornblath_computational_2021}, and human imaging confirms that atrophy patterns follow connectome organization~\cite{dagher_testing_2018}. Epidemic-spreading models incorporating gene expression identify the substantia nigra as the disease epicenter and implicate SNCA and GBA transcription as modulators of vulnerability~\cite{zheng_local_2019}. Longitudinal imaging further shows that atrophy progression is shaped jointly by connectivity, gene expression, and cell-type composition~\cite{tremblay_brain_2021}. Genetic factors also modulate propagation: LRRK2 kinase inhibition can reverse mutation-driven alterations in tau and $\alpha$-synuclein spreading~\cite{lubben_lrrk2_2024}, and glucocerebrosidase activity influences neuronal susceptibility to $\alpha$-synuclein pathology~\cite{henderson_glucocerebrosidase_2020}.
More recent work has also applied graph convolutional networks to learn complex propagation patterns directly from imaging data~\cite{dan_tauflownet_2023}.

Together, these findings provide computational support for the hypothesis that misfolded proteins propagate through brain networks, complementing experimental evidence from cell and animal models, in Alzheimer's and Parkinson's disease.

\subsection{Clearance Mechanisms and Vascular Interactions}

Proteopathic evolution does not depend solely on production and spreading, it also reflects impaired clearance. An imbalance between toxic protein production and clearance is thought to play a major role in the pathogenesis of neurodegenerative disease~\cite{tarasoff2015clearance,harrison2020impaired}. This principle has been formalized in mathematical models: Smoluchowski models identify a \textit{critical clearance} bifurcation point above which amyloid aggregation is suppressed~\cite{meislclearance2020}; extended Smoluchowski models provide a proof of concept for therapies that enhance clearance~\cite{brennan2025network}; and quantitative systems pharmacology models are increasingly being used to translate these ideas into clinical applications.

Building on the Fisher-KPP network framework, models can be coupled with a clearance equation to capture the interplay between protein aggregation and clearance impairment at the brain scale~\cite{brennan_role_2024, brennan2025heterogeneity}. This extension provides computational evidence that deficits in clearance not only exacerbate aggregation locally, but also accelerate and steer the spread of toxic proteins across the connectome.

Clearance pathways---such as cell intrinsic clearance by macroautophagy and non-cell autonomous mechanisms like glial clearance or glymphatic flow---are increasingly recognized as essential to maintaining protein homeostasis, but their precise mechanisms remain debated. Among these, most recent computational models have focused specifically on glymphatic and perivascular transport, suggesting that reduced advection in sleep-deprived but otherwise healthy individuals underlies decreased glymphatic clearance~\cite{vinje_human_2023}, while other studies demonstrate that perivascular transport is far more efficient than diffusion and that its breakdown leads to amyloid accumulation~\cite{diem_simulation_2016}.
A central challenge is to quantify clearance rates \textit{in vivo}, where tracer-based MRI and PET provide valuable indirect estimates. Recent modelling work has begun to bridge this gap by estimating regional glymphatic clearance rates~\cite{brennan2025heterogeneity, vinje_human_2023}, demonstrating how computational approaches can complement experiments and reveal spatial variability in clearance capacity.

Importantly, toxic protein accumulation overwhelms transport, cellular respones, and glial activity~\cite{bennett2018tau, carrillo2014amyloid, canobbio2015role,michalicova2020tau}. This interplay between protein burden and immune response forms the basis of the next key driver of Alzheimer’s pathology that we will discuss.

\subsection{Glial Cells and Neuroinflammation}

Microglia, the brain’s resident immune cells, facilitate clearance of amyloid and tau through proteases and chaperones~\cite{ries2016mechanisms}. Initially protective, they act as phagocytes that monitor the environment and remove toxic proteins. Once overwhelmed, however, they become dysfunctional, exacerbating neurotoxity of the accumulating aggregates by pruning synapses, releasing neurotoxic cytokines, and amplifying neuronal injury and the spread of pathology~\cite{hansen2018microglia}. 
In Alzheimer's disease, for example, tau and amyloid-microglia interactions are described as a double-edged sword \cite{hickman2018microglia}, with neuroinflammation both supporting clearance and accelerating degeneration once a threshold is crossed. 

Recent mathematical models have begun to formalize these interactions~\cite{hao2016mathematical,chamberland_computational_2024,proctor2013}. For example, Chamberland \textit{et al.} (2024) developed a system of differential equations that couples protein burden, microglial activation, and macrophage recruitment, providing a framework to study how inflammation shapes local pathology~\cite{chamberland_computational_2024}. In this model, microglial activation is coupled to toxic protein aggregation through promoting nucleation and impacting constant clearance rates. Although large in scope—capturing the progression of amyloid-$\beta$, tau proteins, neurons, activated astrocytes, microglia, macrophages, and cytokines—the model is not yet validated by experimental data~\cite{chamberland_computational_2024}. To capture microglial responses more tractably, many approaches adopt the common dichotomy of pro-inflammatory $M_1$ versus anti-inflammatory $M_2$ phenotypes and describe their interactions with toxic proteins using coupled differential equations ~\cite{hao2016mathematical,chamberland_computational_2024}.

\subsection{Limitations of Modeling Disease Progression and Mechanisms in Isolation}

Together, these modeling efforts offer essential insight into the biological systems that govern how neurodegeneration emerges and unfolds. They shift focus from the consequences of disease to the mechanisms that drive it. But while they reveal how pathology forms, spreads, and interacts with clearance, glia, vasculature, and metabolism, they often neglect a crucial factor: neuronal activity. As illustrated in Figure~\ref{fig:Overview}B, experimental studies have shown that activity influences nearly every component of disease progression. (1) It accelerates prion-like spreading of misfolded proteins by promoting their release and trans-synaptic transfer~\cite{pooler2013physiological, wu2016neuronal, wu_neuronal_2020}. (2) It regulates glymphatic clearance, where sleep and neuronal firing patterns modulate waste removal efficiency~\cite{xie2013sleep, jiang2024neuronal, murdock2024multisensory}. (3) It drives metabolic demand and mitochondrial stress, linking heightened excitability to increased vulnerability and oxidative burden~\cite{mosconi_brain_2008, sun_restoring_2025}. (4) It modulates microglial activity, where neuronal firing and NMDA receptor activation release ATP that recruits and activates microglia~\cite{dissing-olesen_activation_2014, badimon_negative_2020, umpierre_microglial_2020}. In this sense, activity is not a passive observer of pathology but an active contributor to it. The models reviewed in this section do not account for this causal influence. The next section turns to the open research questions that arise from this gap, and considers how confronting bidirectional interactions between activity and pathology demands new theoretical approaches.

\section{Toward Integrated Models of Neurodegeneration}

Understanding how neurodegeneration alters brain function has motivated two main lines of computational work. The first models changes in neuronal dynamics—shifts in excitability, oscillations, or functional connectivity—that emerge in patients. These models, described in Section II, capture how activity is perturbed by pathology progression. The second, reviewed in Section III, focuses on biological processes underlying progression: protein spreading and interactions with glia, vasculature, and metabolism. Both approaches provide valuable insights but share a limitation: most assume causality flows only from pathology to brain activity. A growing body of evidence shows this assumption is false.

Neuronal activity does not merely reflect disease; it shapes it. Activity modulates protein release, accelerates spreading, alters glial function, and drives metabolic stress. Ignoring this feedback loop creates a blind spot: it disconnects the processes generating pathology from those sustaining cognition. To close this gap, we must rethink how we model the interplay between dynamics and disease. A small but growing set of studies now address this by explicitly modeling bidirectional coupling between neuronal activity and disease. These span scales and mechanisms, from activity-dependent spread to frameworks linking molecular, vascular, and metabolic pathways.

\begin{figure*}[!t]
  \centering
  \includegraphics[width=\linewidth]{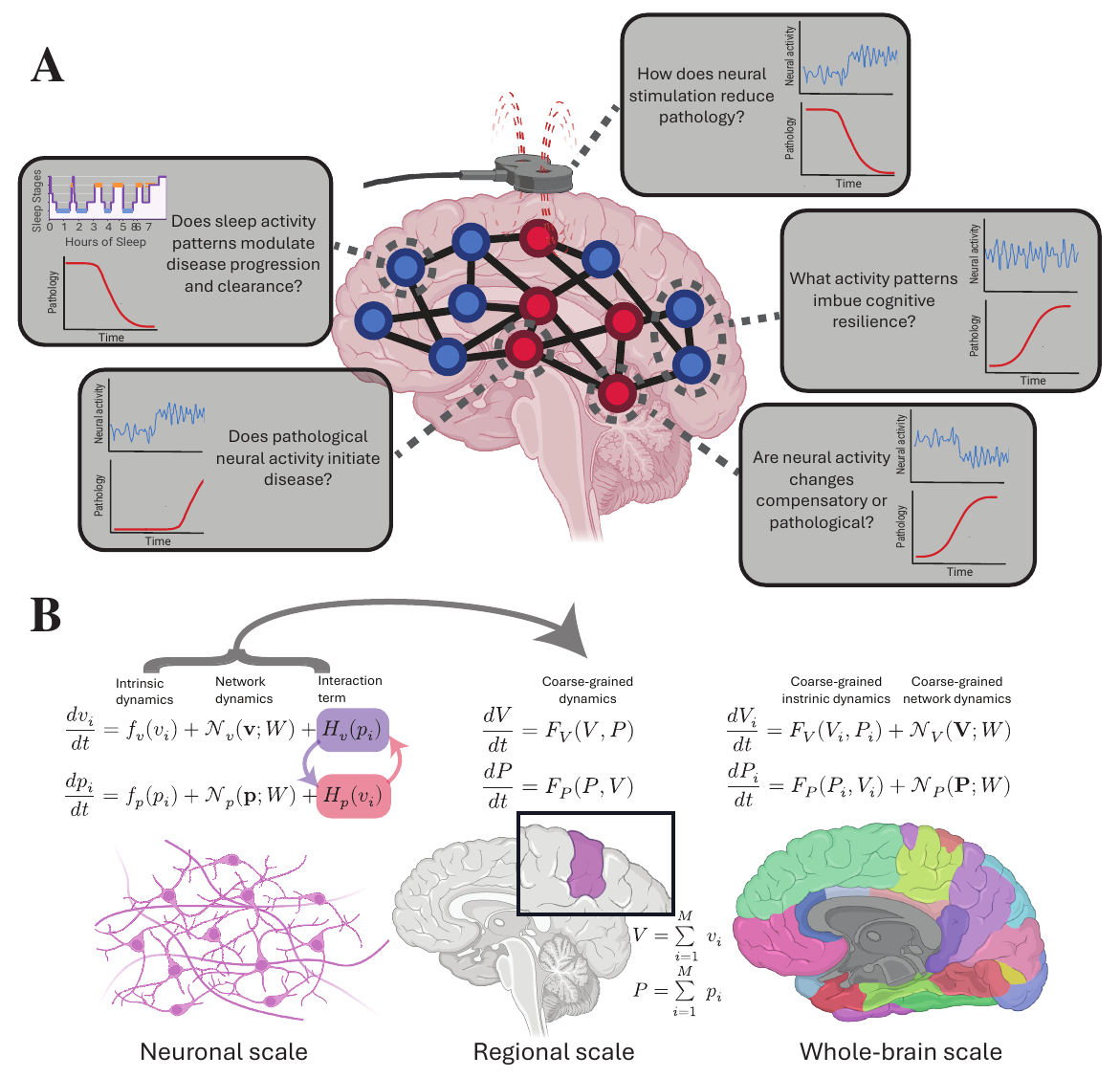}
  \caption{\textbf{Integrated activity–pathology modeling and key scientific questions.}
\textbf{(A)} A schematic for scientific questions that lie at the intersection of neuronal dynamics and pathology.
\textbf{(B)} Conceptual multiscale framework linking neuronal, regional, and whole-brain levels. At each scale, intrinsic and network dynamics are represented, with a bidirectional coupling term between activity and pathology. The feedback loop between pathology and activity is ideally realized on the neuronal scale. 
}
  \label{fig:Integrated_models}
\end{figure*}

\subsection{Existing work on integrated modeling}

Medina-Iturria \textit{et al.} (2017)~\cite{iturria-medina_multifactorial_2017} developed a computational framework accounting for the direct interactions among several imaging-derived biological factors (e.g., amyloid-$\beta$, tau, functional activity indicators, glucose metabolism, cerebrovascular flow, atrophy), their intra-brain spreading across structural and vascular connectomes, and the effects of external inputs (e.g., medication, physical exercise). This multifactorial causal model used a dynamic linear system formalism, in combination with control theory, to study multifactorial mechanisms in neurodegenerative diseases and to indentify single-target and combinatorial personalized therapeutic needs aiming to predict the individual predisposition to respond to different treatments~\cite{iturria-medina_multimodal_2018, lenglos_multivariate_2022}. This framework was extended to incorporate brain-wide distributions of gene expression~\cite{adewale_integrated_2021, adewale_patient-centered_2025} and neuroreceptor densities~\cite{khan_personalized_2022, khan_patient-specific_2023}. Essentially, the extended frameworks modeled the inter-factor interactions (e.g., how local tau deposition impacts functional activity, and vice versa) as a direct function controlled by local genes or receptors, thereby identifying specific molecular mechanisms modulating the disease's multifactorial progression as well as potential drug candidates~\cite{adewale_patient-centered_2025}.

Using a whole-brain modeling approach, De Haan 
\textit{et al.} (2012)~\cite{haan_activity_2012} investigated activity-dependent degeneration with the modeling assumption that regions with higher neuronal activity experience a higher rate of synaptic loss. This was motivated by the observation that functional hub regions in E/MEG connectivity studies are more prone to amyloid deposition and pathology, and that excessive neuronal activity leads to higher amyloid-$\beta$ deposition. Surprisingly, however, simulations first showed a transient increase in neuronal activity, particularly in hub regions, before subsequent decrease in activity. The authors argued that this transient hyperactivity may be caused by disinhibition due to synaptic degradation of pre-synaptic inhibitory neurons. This study highlights that simple, local feedback rules between activity and pathology can explain disease-related phenomena, such as early-stage hyperactivity and late-stage hypoactivity observed in Alzheimer's animal models.  
Different theurapeutic strategies were explored computationally using the same whole-brain model with activity-dependent degeneration, predicting the stimulation of excitatory neurons as the most promisting target~\cite{haan_altering_2017}. However, in this modeling approach, the causal interactions between disease and neuronal activity is included implicitly in the neuronal dynamics model, and does not capture the impact that neuronal activity has on the disease progression itself. 

Other approaches have investigated the coupled dynamics of neuronal activity and disease progression over time, motivated by evidence that transneuronal spreading of protein pathology is accelerated by neuronal firing~\cite{pooler2013physiological, wu2016neuronal}.
Alexandersen \textit{et al.} (2024)~\cite{alexandersen_neuronal_2024} developed a mathematical model for such co-evolution of neuronal activity and disease progression, where neuronal firing increases pathological protein spread and the pathological proteins damage neuronal firing dynamics. This model demonstrated that neuronal activity may not only affect disease progression patterns, but also initiate disease onset. In particular, the model proposed a theory in which gradients of neuronal activity determine where disease first emerges, with regions (or neurons) at the lower end of the gradient being the most likely disease epicenters. This theory was then tested by incorporating FDG and amyloid-$\beta$ PET neuroimaging data into the model, under the assumption that amyloid-$\beta$ causes hyperactivity. The model correctly recovered the entorhinal cortex as the initial tau epicenter. Across individuals, stronger predicted entorhinal seeding corresponded to higher empirical entorhinal tau, suggesting that brain-wide activity patterns may bias the site of tau onset in Alzheimer's disease~\cite{alexandersen_neuronal_2025}.
Cabrera-Alvarez \textit{et al.} (2024)~\cite{cabrera-alvarez_multiscale_2024} simulated the co-evolution of amyloid-$\beta$ and tau pathology together with neuronal activity on the whole-brain scale, where hyperactivity leads to higher productions of A$\beta$ and a biased prion-like transport of tau to high-activity regions, finding that changes in inhibition due to A$\beta$ captures neurophysiological changes observed in Alzheimer's. 

Together, these integrated modeling efforts demonstrate that it is both feasible and informative to couple neuronal dynamics with disease biology. But they also make it clear that much remains unexplored: the space of possible interactions is vast, and many clinically relevant problems lie outside the scope of current models. This motivates a closer look at the kinds of scientific questions that can only be addressed when both domains are modeled together.

\subsection{Scientific Questions That Call for Integrated Modeling}

While models of neuronal activity and disease processes each provide mechanistic insight, the most important questions in neurodegeneration arise at their intersection. These cannot be answered by considering activity or pathology alone, as they depend on how the two systems interact over time. Below, we highlight several such questions (Figure~\ref{fig:Integrated_models}A) and explain why each requires an integrated, bidirectionally coupled modeling approach.

\subsubsection{\textbf{What determines where and when pathology first emerges?}}
The entorhinal cortex is widely believed to be the epicenter of tau pathology in Alzheimer's disease and primary age-related tauopathy. However, the reason for this vulnerability is unknown. Neuronal activity has long been implicated in the genesis of neurodegeneration, and several recent studies suggest that regional activity levels play a crucial role in the progression of tau pathology from the entorhinal cortex to the rest of the brain, a key event in Alzheimer's progression~\cite{giorgio_amyloid_2024, hojjati_inter-network_2025, roemer-cassiano_amyloid-associated_2025}. A recent mathematical model has begun to address this question, showing that neuronal activity can even initiate tau pathology and correctly predicting the entorhinal cortex as the most vulnerable epicenter~\cite{alexandersen_neuronal_2024, alexandersen_neuronal_2025}. A growing body of evidence suggests that neuronal activity—particularly whole-brain patterns of regional activity—plays a decisive role in the initiation of neurodegeneration. Specific activity patterns may contribute to ageing and disease, raising the possibility that targeted neuronal stimulation could not only mitigate symptoms but also prevent disease onset by reshaping brain-wide dynamics. Given this interplay between activity and pathogenesis, mathematical models are well positioned to generate theories and hypotheses on how neurodegenerative disease begins and how its initiation might be halted. 

\subsubsection{\textbf{Are observed changes in activity compensatory or degenerative?}}

Changes in neuronal activity during the early stages of Alzheimer’s disease can reflect either beneficial adaptation or detrimental dysfunction. For example, hippocampal hyperactivity in individuals with mild cognitive impairment has been associated with better short-term memory performance in some studies~\cite{kircher_hippocampal_2007}, but also with accelerated tau accumulation in connected cortical regions~\cite{huijbers_tau_2019}. Similarly, increased network synchrony has been interpreted in some cases as a compensatory mechanism that supports cognitive performance despite pathology~\cite{gaubert_eeg_2019}, while in other contexts it is linked to pathological hypersynchrony that contributes to network instability and cognitive decline~\cite{verret_inhibitory_2012}. This lack of consensus underscores the difficulty in determining whether suppressing hyperactivity or hypersynchrony would remove a harmful driver or an adaptive response. Bidirectionally coupled models that track both neuronal activity and pathology over time are uniquely positioned to evaluate such counterfactual scenarios and predict whether observed changes are ultimately protective, harmful, or both.

\subsubsection{\textbf{How and when does neuronal stimulation reduce protein burden?}}
Non-invasive stimulation protocols, such as gamma-frequency sensory stimulation, have been shown to reduce amyloid-$\beta$ and tau burden in mouse models~\cite{martorell_multi-sensory_2019, iaccarino_gamma_2016}, potentially via enhanced microglial clearance~\cite{iaccarino_gamma_2016} and altered neuronal firing patterns~\cite{adaikkan_gamma_2019}. Early human studies suggest that similar approaches can modulate biomarkers and improve certain cognitive measures~\cite{chan_gamma_2022}, but the durability and generalizability of these effects remain unclear. Because stimulation alters network dynamics on short timescales while protein aggregation and clearance occur over much longer periods, the same intervention could yield divergent outcomes depending on dose, timing, and disease stage. Models that integrate the acute effects of stimulation on neuronal activity with the slower dynamics of pathology progression are essential for predicting these long-term outcomes and identifying regimes where benefits outweigh potential risks.

\subsubsection{\textbf{Why do some individuals maintain cognitive function despite high pathology?}}
Some individuals remain cognitively intact for years despite high amyloid or tau burdens on PET scans, a phenomenon often attributed to “cognitive reserve”~\cite{aizenstein_frequent_2008, rentz_cognition_2010}. Cognitive reserve likely reflects the brain’s ability to dynamically reconfigure networks and maintain performance despite ongoing damage~\cite{steffener_supporting_2011, franzmeier_left_2017}. However, pathology also constrains plasticity by altering connectivity and excitability, meaning reserve is not static but co-evolves with disease~\cite{kocagoncu_tau_2020, terry_physical_1991}. Integrated models can simulate alternative trajectories for individuals with identical pathology loads but different capacities for reorganization, clarifying the conditions under which compensation sustains function versus when it fails. Longitudinal studies show that reserve can delay the onset of clinical symptoms, but once breakdown occurs, decline is often rapid~\cite{hall_education_2007}, suggesting a bifurcation in the co-evolution of disease and neuronal dynamics leading to a delayed and abrupt transition in disease progression.

\subsubsection{\textbf{How do sleep-linked activity patterns change vascular clearance and does boosting slow waves slow the disease?}}

Deep-sleep slow waves co-occur with large-scale neural synchrony, cerebrovascular/CSF pulsations, and enhanced metabolite clearance, suggesting that specific activity patterns can modulate the efficiency of waste removal~\cite{fultz_coupled_2019, hablitz_increased_2019}. Experimental and human evidence further indicates that disrupting sleep or slow-wave activity elevates soluble A$\beta$/tau on short timescales, whereas augmenting slow waves via closed-loop stimulation can enhance oscillations and improve memory~\cite{shokri-kojori_-amyloid_2018, ngo_auditory_2013}. The modeling question is not only whether oscillations increase clearance, but how this coupling co-evolves with disease: accumulating pathology also degrades sleep architecture and impairs neurovascular responsiveness~\cite{mander_-amyloid_2015, nortley_amyloid_2019}. Because neural entrainment acts over seconds-to-minutes, while aggregation and clearance evolve over months–to-years, integrated, bidirectionally coupled models are needed to link slow-wave modulation to long-term pathology and to identify the vascular conditions under which such interventions are likely to succeed.

\subsection{Theoretical and Computational Challenges in Model Integration}

If the previous section argued why integration is necessary, this one explains why it remains difficult. Models spanning neuronal dynamics and disease processes face theoretical and computational challenges. Here, we highlight two key ones: timescale separation and the absence of proper mean-field descriptions of coupled systems.

\subsubsection{Timescale Separation and Multiscale Simulation}

One key challenge in simulating and analyzing multiscale models is the mismatch in timescales across different biological processes. Neuronal dynamics evolve on the order of milliseconds to seconds, whereas protein aggregation, glial responses, and tissue-level degeneration unfold over days, months, or years. Clearance mechanisms and vascular processes may additionally be modulated by circadian rhythms and sleep.

When the fast process—in our case neuronal dynamics—has no causal impact on the other processes, one can simulate the slow processes independently (e.g., prion-like spreading) and adjust the fast activity afterwards. This is no longer possible when the fast process feeds back into the slow process. In that case, the simulation must in principle be resolved at the finest timescale: modelling years of neurodegeneration with second-by-second neural dynamics. This is both computationally prohibitive and conceptually undesirable, as it obscures the system’s distinct temporal structure. Long-term changes may depend not on instantaneous fluctuations but on averaged patterns or cumulative load.

Multiple-timescale theory offers one way to avoid this problem by exploiting timescale separation: the fast subsystem is assumed to equilibrate, or otherwise settle into a regular attractor, on a timescale that is effectively instantaneous relative to the slow variables~\cite{kuehn_multiple_2015}. The slow dynamics are then determined by the average influence of the fast subsystem, and the fast variables respond to the slow variables as slowly varying parameters.
When the fast subsystem has a stable equilibrium for fixed slow variables, this approach is on firm ground. The equilibria form a critical manifold, and results such as Fenichel’s theorem~\cite{fenichel_geometric_1979} guarantee that this manifold persists when the timescale separation is large but finite, allowing a rigorous reduction to a slow flow along it.

Fast neuronal dynamics, however, rarely behave so simply. In some cases, population activity is approximately periodic, as in sustained oscillations. Here, the “critical manifold’’ is a family of stable periodic orbits, and reduction techniques must combine invariant manifold theory with averaging over the fast phase. This combination allows slow variables to evolve according to the mean effect of the oscillations, and the fast dynamics to be modulated slowly in return. Such methods are standard in the analysis of neuronal bursting, where slow ionic processes shape fast spiking rhythms~\cite{cressman_influence_2009, houssaini_epileptor_2020}, but are more involved than the fixed-point case and rely on the existence of a coherent oscillatory regime.

For irregular or aperiodic fast dynamics, there may be no low-dimensional invariant set to serve as a geometric basis for reduction. In such cases, some studies apply stochastic averaging techniques to derive effective dynamics for slow variables, while a recent study developed mean-field reductions that capture the average behavior of irregularly firing neurons~\cite{galtier_multiscale_2012, clusella_exact_2024}.
A common workaround is to describe the fast process statistically—through coarse variables such as firing rate distributions or correlation structure—and then couple these reduced variables to the slow subsystem~\cite{cabrera-alvarez_multiscale_2024, alexandersen_multi-scale_2023}. However, when this is done in a post-hoc manner, with each subsystem reduced in isolation before coupling, important aspects of the joint microscale dynamics may be lost. Capturing how fast and slow processes influence each other at the microscopic level may require new analytical tools capable of reducing the coupled dynamics directly, rather than combining separate reductions after the fact. This limitation motivates the next section.

\subsubsection{Toward Mean-Field Descriptions of Interacting Processes}

Mean-field modelling refers to bottom-up approaches for reducing complex microscopic systems to lower-dimensional descriptions, where a large system of interacting components is instead described by higher-order variables, such as averages. In neuroscience, for example, spiking neuron networks can be approximated by neural mass or neural field models. Traditional neural mass models, such as the Jansen-Rit~\cite{jansen_electroencephalogram_1995} and Wilson-Cowan models~\cite{wilson_excitatory_1972}, are often approximations of average population firing rates. However, new  theoretical developments such as the Ott-Antonsen ansatz~\cite{ott_low_2008} has ushered in a new wave of exact mean-field reductions (see Bick \textit{et al.} (2020)~\cite{bick_understanding_2020} for a review), where the average behaviour of infinitely large networks of single neuron models, such as quadratic integrate-and-fire neurons, can be captured exactly by a low-dimensional dynamical system~\cite{montbrio_macroscopic_2015, el_boustani_master_2009, byrne_next-generation_2020}. In protein aggregation studies, detailed molecular kinetics can be reduced to coarse equations for regional concentrations. These reduced models capture the average behaviour of large populations and are widely used to connect microscale mechanisms to mesoscopic or macroscopic dynamics.

In current practice, such mean-field reductions are usually performed in isolation for each process—neuronal dynamics, protein spreading, glial or vascular changes, and the resulting macroscopic variables are then linked afterwards in a “post-hoc” coupling~\cite{alexandersen_multi-scale_2023, cabrera-alvarez_multiscale_2024}. While this is convenient, it risks omitting important features of the microscale interactions between processes. For example, synaptic activity may influence protein release at the level of individual synapses, or local protein accumulation may alter neuronal excitability cell-by-cell. If each process is reduced independently before coupling, these fine-scale feedbacks may be lost or distorted in the macroscopic model.

A more principled approach would be to develop joint mean-field theories in which the reduction is applied to the coupled microscopic dynamics. In the context of neurodegeneration, such a microscopic model might include prion-like protein spreading and clearance, vascular or glial state changes, and recurrent neuronal networks generating oscillations or irregular spiking. The mean-field reduction would then yield a mesoscopic model in which all these components co-evolve according to interactions inherited directly from the microscale, with Figure~\ref{fig:Integrated_models}B providing a generic example of such a model. 
The outcome would be a closed set of equations where the coupling between neural and pathological processes is consistent with their joint microscopic origins.

While such joint mean-field reductions are conceptually straightforward, they remain largely unexplored, particularly for heterogeneous, multi-component systems relevant to neurodegeneration. Developing these approaches would not only improve model accuracy, but could also reveal undiscovered phenomena and avenues for treatment. For example, noninvasive neuronal stimulation has shown promising results in not only ameliorating cognitive symptoms, but also reversing pathology in neurodegenerative diseases~\cite{martorell_multi-sensory_2019, iaccarino_gamma_2016, suk_vibrotactile_2023}. How these stimulation treatments work remains unknown but likely involves changes in neuronal synchrony, plasticity, prion-like spreading, and clearance. Current modeling efforts are not equipped to provide causal accounts for the effects of neuronal stimulation, and joint mean-field models of neuronal processes will fill this gap. 


\subsection{Integrating Causal and Data-Driven Models}
Mechanistic models rely on \textit{a priori} biological knowledge, whereas empirical models use data-driven approaches to characterize and predict disease without strong assumptions (see reviews~\cite{khan_beyond_2024, young_data-driven_2024}). Their flexibility enables integration of heterogeneous data across biological scales, supporting applications ranging from mapping disease trajectories to individualized patient stratification. For example, empirical models integrate post-mortem brain omics (epigenomics, transcriptomics, proteomics, metabolomics) with neuroimaging and blood-based measures to stratify the AD spectrum at molecular resolution and translate post-mortem findings to living individuals~\cite{iturria-medina_unified_2022, iturria-medina_translating_2025}. Integrating mechanistic and empirical models, leveraging complementary strengths, promises more accurate predictions and richer biological understanding of neurodegenerative disease.

\section{Conclusion}
Computational models are indispensable for understanding neurodegenerative disease. Over the past decade, distinct traditions have emerged: some focus on neuronal dynamics and circuit dysfunction, others on biological processes driving progression, such as protein aggregation, glial responses, vascular impairment, and clearance failure. Each has yielded valuable insights, but mostly in isolation. Growing evidence shows these mechanisms are deeply interconnected and cannot be understood separately. Neuronal activity shapes processes such as protein release and clearance, while accumulating pathology feeds back to disrupt network dynamics.

To address fundamental questions—where and why pathology emerges, how resilience arises, and how interventions alter progression—we must develop integrative models that capture this feedback. Bridging neuronal dynamics with disease biology is no longer a theoretical ambition but a practical requirement for interpreting multimodal data and designing interventions that are both targeted and effective. The road ahead will demand theoretical advances and flexible multiscale frameworks capable of capturing interactions across biological and dynamical domains. Progress in this direction will deepen our understanding of how these processes shape one another and clarify the mechanisms through which interventions exert their effects. By embedding neuronal activity within its biological context, integrated models can serve as a crtical bridge between empirical data and therapeutic design, advancing both scientific insight and translational relevance.


\bibliographystyle{IEEEtran}
\bibliography{IEEEabrv,refs.bib}

\end{document}